\documentstyle[prb,aps,epsfig,floats]{revtex}

\begin{document}
\draft
\preprint{}
%
%following two lines are for two column format
\twocolumn[\hsize\textwidth\columnwidth\hsize\csname
@twocolumnfalse\endcsname
%
%This will double space the paper (should also remove two column format)
%\def\@doubleleading{1.6}
%\def\baselinestretch{\@doubleleading}
%\def\doublespace{\def\baselinestretch{\@doubleleading}\normalsize}
%\doublespace

\title{Universality of the Electronic 
Structure from a Half Filled CuO$_2$ Plane}

\author{F. Ronning\cite{fr}, C. Kim\cite{cyk}, K.M. Shen, 
N.P. Armitage\cite{npa}, A. Damascelli,
D.H. Lu, D.L. Feng\cite{dlf}, and Z.-X. Shen}
\address{
Department of Physics, Applied Physics and Stanford Synchrotron
Radiation Laboratory,\\ Stanford University, Stanford, CA 94305,
USA}
\author{L.L. Miller}
\address{Department of Physics, Iowa State University, Ames Iowa, 50011}
\author{Y.-J. Kim, F. Chou}
\address{Department of Physics, Massachusetts Institute of Technology, 
Cambridge, MA, 02139}
\author{I. Terasaki}
\address{Department of Applied Physics, Waseda University, Tokyo 
169-8555, Japan}

\date{\today}
\maketitle
\begin{abstract}

We present angle resolved photoemission (ARPES) data from insulating
A$_2$CuO$_2$X$_2$ (A=Sr, Ca; X=Cl, Br), Sr$_2$Cu$_3$O$_4$Cl$_2$, and
Bi$_2$Sr$_2$MCu$_2$O$_{8}$ (M=Er, Dy) single crystals which illustrate
that the low energy electronic structure of the half-filled cuprates are
independent of the apical atom.  By performing a photon energy dependent
study on Ca$_2$CuO$_2$Cl$_2$ we are able to distinguish between features
which are intrinsic and those which are a result of the photoemission
matrix elements.  We find that the dispersion is independent of photon
energy, while an asymmetry in the electron occupation probability across
the antiferromagnetic zone boundary is robust to variations in photon
energy.  Finally, we show that the {\it d}-wave-like dispersion which
exists in the insulator along the antiferromagnetic zone boundary does not
precisely fit the simple {\it d}-wave functional form near the nodal
direction.

\end{abstract}
% \pacs{PACS numbers: 74.72.Jb, 74.72.Jt, 79.60.Bm}
%
\vskip2pc]
\narrowtext

\section{Introduction}

One approach to unraveling the mysteries of the high T$_{c}$ 
superconductors is to understand how the electronic structure evolves 
from an antiferromagnetic insulator to a superconductor upon doping.  
The first step is then to understand the insulator to metal 
transition.  Although Mott qualitatively described how a material, 
predicted by band theory to be a metal, would in fact be an 
insulator,\cite{Mott49} it remains unclear as to how the details of 
the electronic structure evolve from a half-filled metal to a Mott 
insulator.  Questions such as how does the gap open, and what happens to 
the information contained in the occupation probability n(${\bf k}$) 
when strong correlations drive the system insulating need to be 
answered.  The technique of angle resolved photoemission (ARPES) is a 
natural choice to do this.

Of course, with every well posed question, there are experimental
challenges which make it difficult to answer these questions.  
La$_{2-x}$Sr$_{x}$CuO$_{4}$(LSCO) is one system which has been
successfully grown from half filling to heavily overdoped; however, ARPES
results on these crystals remain somewhat uncertain as illustrated by the
extremely broad spectra of La$_{2}$CuO$_{4}$.\cite{InoPRB00}
Nd$_{2-x}$Ce$_{x}$CuO$_{4}$(NCCO) is another system which spans a similar
doping range, and has attempted to answer the above
questions;\cite{ArmitageNCO,HarimaPRB01} however, it is not clear whether
the results can be generalized to include the hole doped cuprates.  
YBa$_2$Cu$_3$O$_{6+\delta}$(YBCO), which can cover the range of
half-filling to optimal doping, might appear as the next suitable
candidate, but the presence of a surface state and the existence of one
dimensional chains has clouded the interpretation of its bulk electronic
structure.\cite{SchabelPRB98} The
Bi$_2$Sr$_2$CaCu$_2$O$_{8+\delta}$(Bi2212) system, with its extremely good
cleavage plane, has been ideal for ARPES. For precisely this reason, the
majority of ARPES data on the high T$_{c}$'s to date have come from
Bi2212.  Unfortunately, high quality Bi2212 crystals at very low dopings
have not been achieved.  On the other hand, A$_2$CuO$_2$Cl$_2$ (A=Sr,Ca)
also cleaves extremely well and gives high quality ARPES data comparable
to that seen in optimally doped Bi2212, but in this case, single crystals
have only been available at half filling.  As a result, there exists no
perfect system to study the electronic structure from the
antiferromagnetic insulator to the heavily over doped metal in the hole
doped cuprates.

The solution has been to make the reasonable assumption that the low
energy physics of the Bi2212 and A$_2$CuO$_2$Cl$_2$(ACOC) systems are
identical due to the fact that the CuO$_{2}$ planes are common to both
structures.  Thus the entire doping range can be studied.  However, these
two systems also have several differences.  Sr$_2$CuO$_2$Cl$_2$(SCOC) and
Ca$_2$CuO$_2$Cl$_2$(CCOC) have a Cu-O-Cu distance of 3.97\AA\ and 3.87\AA\
respectively,\cite{Grande75} compared to 3.83\AA\ for
Bi2212,\cite{LynchOlsonBook} and they also do not possess the orthorhombic
distortion and superstructure effects which plague Bi2212.  Finally, the
most striking difference in ACOC is that the apical oxygen has been
replaced with a chlorine atom.  It is important to test whether or not
these differences can have an effect on the low energy electronic
structure.

In this paper we will validate the assumption that low energy ARPES data
on the oxyhalides are indeed representative of photoemission from a
generic, half-filled CuO$_{2}$ plane, and thus may be reasonably compared
with ARPES data on hole doped Bi2212.  This will be done by showing that
replacing the apical chlorine with the larger, and less electronegative
bromine has no effect on the low energy electronic structure.  
Furthermore, it will also be shown that heavily underdoped Bi2212 near
half filling, despite having relatively poor spectral quality, is
qualitatively consistent with the results on the oxyhalides.  Finally,
ARPES on Sr$_2$Cu$_3$O$_4$Cl$_2$, which contains an additional copper atom
in every other CuO$_{2}$ plaquette, demonstrates that the lowest lying
excitations attributed to a Zhang-Rice singlet\cite{ZhangRice88} is
surprisingly unaffected by even a seemingly large modification of the
CuO$_{2}$ plane.

Having shown that the ARPES results of the half filled Mott insulating
cuprate are not system dependent, we will turn our attention in the latter
half to extracting the physics of the half filled CuO$_{2}$ plane
contained in the single particle spectral function, A(${\bf k}, \omega$).  
Because matrix elements modulate the measured photoemission intensity, it
is important to distinguish which features of the data are intrinsic and
which are extrinsic.  However, photon energy ($E_{\gamma}$)
dependent studies designed specifically to test the influence of the
matrix element on both the dispersion $E({\bf 
k})$,\cite{HaffnerPRB00,DurrPRB00} and the intensity n(${\bf
k}$),\cite{HaffnerPRB01} have had conflicting reports on the magnitude of
variations caused by matrix elements.  So to test whether or not the
dispersion and the ${\bf k}$ dependence of the spectral weight are
impressive manifestations of matrix element modulations, we have performed
ARPES n(${\bf k}$) mappings over the entire Brillouin zone for 5 different
photon energies, and examined the (0,0) to $(\pi,\pi)$ cut for 13 photon
energies.  We find that the dispersion is independent of photon energy as
one might expect, and that with few exceptions, the remnant Fermi
surface\cite{RonningScience98} is robust despite observing strong
variations in spectral weight caused by matrix elements.

With an understanding of the matrix element we can finally turn our
attention to the physics.  Specifically, we will focus on the
$d$-wave-like dispersion found in the insulator.\cite{RonningScience98}
This observation allows for a natural connection between the $d$-wave form
of the high energy pseudogap seen in underdoped Bi2212 and the dispersion
of the insulator as first suggested by Laughlin.\cite{LaughlinPRL97} This
connection is particularly intriguing in light of the fact that the high
energy pseudogap and low energy pseudogap appear to be
correlated.\cite{WhitePRB96} The latter of which is directly related to
the superconducting gap, thus linking antiferromagnetism which is
responsible for the dispersion of the insulator to $d$-wave
superconductivity.

The original ARPES dispersion data on CCOC left some ambiguity as to
whether or not the detailed dispersion near the node exactly fit the
$d$-wave functional form.  A linear dispersion at the node of the form
$E(\pi/2,\pi/2)-E({\bf k})\propto ||{\bf k} - (\pi/2,\pi/2)||$ is highly
nontrivial and several theories which attempt to connect the insulator to
the superconductor predict precisely such a discontinuity in the
derivative of the dispersion at the node
\cite{LaughlinPRL97,WengPRB01,RabelloPRL98}.  However, the $t$-$J$ model
with next nearest neighbor hopping terms $t^{\prime}$ and
$t^{\prime\prime}$, which correctly describes gross features of the
dispersion, has a functional form of $cos2{k_{x}a}+cos2{k_{y}a}$ which is
analytic at $k_{x}$=$k_{y}$.  To investigate this issue we performed ultra
high resolution ARPES experiments along the antiferromagnetic Brillouin
zone(AFBZ) to determine the exact nature of the dispersion near
$k_{x}$=$k_{y}$.  From this data we find that the dispersion near the node
is non-linear and thus can not be fit by the simple $d$-wave functional
form of $|cosk_{x}a-cosk_{y}a|$.

The paper is organized as follows.  Section II presents the experimental
details.  Sections III through VI present photoemission results from the
various insulating parent cuprates illustrating that the low energy
spectra is independent of the apical site.  Section VII is a photon energy
dependent investigation of CCOC.  Section VIII compares the dispersion of
CCOC with the pure d-wave functional form.  Section IX discusses the
experimental results with regards to current theoretical understanding
after which we conclude.

\section{Experimental}

A$_2$CuO$_2$X$_2$ (A=Sr, Ca; X=Cl, Br), Sr$_2$Cu$_3$O$_4$Cl$_2$, and
Bi$_2$Sr$_2$MCu$_2$O$_{8}$ (M=Er, Dy)  single crystals were grown by a
slow cool flux method.\cite{LancePRB90,KitajimaJPCM99} ARPES experiments
were performed at beamlines V-3 and V-4 of the Stanford Synchrotron
Radiation Laboratory.  The beamline V-4 system is capable of achieving an
energy and angular resolution better than 15meV and 0.25$^{\circ}$.  
However, for the majority of this work such high resolution is not
necessary due to the extremely broad features which are being studied.  
Thus, the high resolution is only occasionally utilized while the
remaining data presented uses a resolution of $\leq$70meV and $\pm
1^{\circ}$.  Crystals were oriented prior to the experiment by Laue back
reflection, and cleaved {\it in situ} at a base pressure better than 5
$\times$ 10$^{-11}$ torr.  The photon flux was varied to determine whether
or not the insulating sample was charging. If it was then the sample
temperature was raised when possible to eliminate this effect. A
lineshape which is independent of the photon flux is taken as proof that
there is no charging despite the fact that the samples are insulators.
Slight charging is observed for the Ca$_2$CuO$_2$Br$_2$ and
Sr$_2$Cu$_3$O$_4$Cl$_2$ samples at room temperature, but the results were
reproducible and the spectra simply shifted to higher binding energy with
increased flux indicating a uniform potential barrier formed due to
charging.

We also note that the minimum binding energy of an insulator can vary up
to 1eV from cleave to cleave even in the absence of any charging. The
reason for this is not known with certainty, but is likely due to
different pinning sites for different cleaves. For the oxychlorides we
typically find the centroid of the minimum binding energy feature to lie
between 0.5 and 0.8eV below the chemical potential which is determined by
the $E_{F}$ of a reference gold sample in electrical contact.

\section{S\lowercase{r$_2$}C\lowercase{u}O\lowercase{$_2$}C\lowercase{l$_2$} 
and C\lowercase{a$_2$}C\lowercase{u}O\lowercase{$_2$}C\lowercase{l$_2$}}

\begin{figure}[tbp]
	\centering
\includegraphics[width=3in]{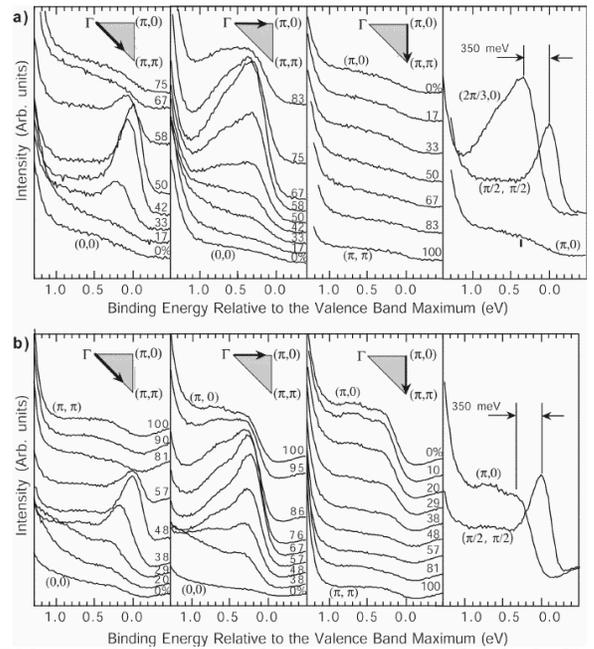} 
\caption[ARPES spectra of 
Sr$_2$CuO$_2$Cl$_2$ and Ca$_2$CuO$_2$Cl$_2$]{EDCs of 
Sr$_2$CuO$_2$Cl$_2$(a) and Ca$_2$CuO$_2$Cl$_2$(b) along the high 
symmetry direction indicated in the insets.  The final panel indicates 
the magnitude of the $d$-wave-like dispersion seen in the oxyhalides, 
and the difference in lineshape exhibited as a function of ${\bf k}$.  The 
measurement conditions were T=150K, $E_{\gamma}$=22.4eV and T=100K, 
$E_{\gamma}$=25.2eV for Sr$_2$CuO$_2$Cl$_2$ and Ca$_2$CuO$_2$Cl$_2$, 
respectively.}
	\label{SCOCvsCCOC}
\end{figure}

Figure \ref{SCOCvsCCOC} presents a comparison between SCOC and CCOC along
the high symmetry directions.  The two are nearly identical.  This is to
be expected as Sr and Ca are isovalent and lie between the CuO$_{2}$
planes.  From (0,0) to $(\pi,\pi)$ they show a feature which emerges from
the background, disperses towards the chemical potential, reaches a
maximum at $(\pi/2,\pi/2)$ and then loses weight rapidly as it pushes back
to higher binding energy.\cite{WellsPRL95} We note that the centroid of
the lowest energy excitation at $(\pi/2,\pi/2)$ still lies well below the
chemical potential (off scale), indicative of the fact that these crystals
are Mott insulators.  Along the (0,0) to $(\pi,0)$ cut a more intense and
significantly more asymmetric peak is observed.  Under certain
experimental conditions it becomes clear that this strong asymmetry is due
to the presence of a second feature which lies approximately 600meV below
the main band.\cite{KimPRL98} The dispersion of this second feature
appears similar to that of the lower binding energy feature, although the
intensity is not.  The feature along the $(\pi,0)$ cut does not show much
dispersion, and lies approximately 350meV below the maximum in dispersion
at $(\pi/2,\pi/2)$. In spectra where the peak positions are not as clear
we will attempt to make a quantitative comparison by taking the minimum of
the second derivative of the spectra. We do not present any detailed fits
of the spectra due to a lack of theoretical understanding of the broad
spectral lineshapes inherent to the half-filled cuprates. We note that the
broad spectral lineshapes of the half-filled cuprates are an intrinsic
properties of the spectra, and not a result of the sample charging. As
stated above, this is confirmed in ACOC by varying the photon flux and
observing no change in the spectra.

The one significant difference between SCOC and CCOC is the
observed spectral weight at $(\pi,0)$ in CCOC which is suppressed in SCOC.
On going from $(\pi,0)$ to $(\pi,\pi)$ it can be seen that this weight
vanishes quickly.  Although it is not clear why the matrix elements would
favor CCOC over SCOC in the $(\pi,0)$ region, it is this difference which
facilitated the identification of a remnant Fermi surface in CCOC which
was not observed in previous studies on SCOC\cite{RonningScience98}. 

One of the most significant features of the insulator is the dispersion
between $(\pi/2,\pi/2)$ and $(\pi,0)$. There are two aspects to notice in
the right most panels of figure \ref{SCOCvsCCOC}.  The most important is
clearly the difference of roughly 350meV in dispersion.  This is the
magnitude of the $d$-wave-like modulation seen in the insulator.  The
second is the strong difference in lineshape.  At $(\pi/2,\pi/2)$ the
spectra show a clear peak which resembles a quasiparticle like peak,
albeit with a {\it very} large width($\sim$300meV), while at $(\pi,0)$,
the spectra merges more continuously into the high energy background,
somewhat resembling a step function.  These will be the benchmarks to
which the other compounds will be compared in determining whether or not
the data from SCOC and CCOC are representative of a single half-filled
CuO$_{2}$ plane.

\section{C\lowercase{a$_2$}C\lowercase{u}O\lowercase{$_2$}B\lowercase{r$_2$}}

\begin{figure}[tb]
	\centering
\includegraphics[width=3in]{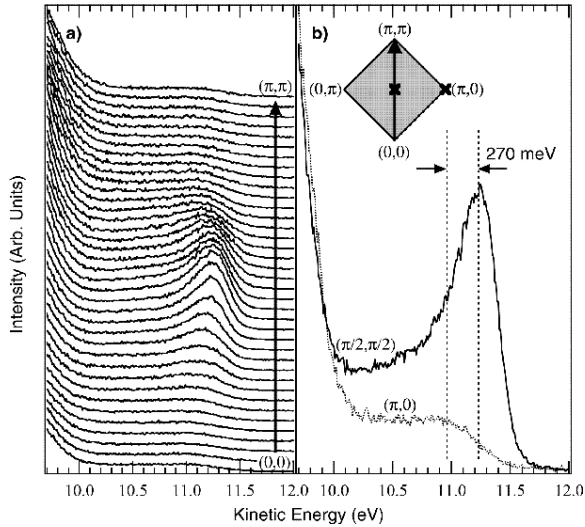} 
\vspace{0.1in}
\caption[ARPES spectra of 
Ca$_2$CuO$_2$Br$_2$]{a) EDCs of Ca$_2$CuO$_2$Br$_2$ from 
(0,0) to $(\pi,\pi)$ taken with 16.5eV photons at T=372K. b) 
Comparison of $(\pi/2,\pi/2)$(solid line) and $(\pi,0)$(dotted line). 
The dashed lines mark the positions of maximum curvature.}
	\label{CCOB}
\end{figure}

The biggest difference between ACOC and other high T$_{c}$ cuprates is the
presence of chlorine as opposed to oxygen in the apical site.  To see
whether or not the apical site has an effect on the electronic structure
we present in figure \ref{CCOB} ARPES data from Ca$_2$CuO$_2$Br$_2$ where
bromine has replaced chlorine in the apical site.  From band theory
calculations the binding energy of the apical orbital is -3.4eV, -2.8eV,
and -2.6eV for Ca$_2$CuO$_2$Cl$_2$, Ca$_2$CuO$_2$Br$_2$, and La$_2$CuO$_4$ 
respectively,\cite{MattheissPRB90} which suggests that CCOB
is a suitable candidate to determine the effect of the apical chlorine as
compared with an apical oxygen on the low energy electronic structure. On
the cut from (0,0) to $(\pi,\pi)$ a dispersive feature is clearly observed
with a minimum in binding energy at $(\pi/2,\pi/2)$ and an overall
bandwidth of approximately 300 meV. In panel (b) the spectra at
$(\pi/2,\pi/2)$ and $(\pi,0)$ are compared.  These data look almost
identical to those from CCOC which are shown in figure \ref{SCOCvsCCOC}.  
Specifically, the lineshape at $(\pi/2,\pi/2)$ is quite sharp and well
defined compared with the broad feature seen at $(\pi,0)$.  The energy
difference of 270 meV (determined by the minimum of the second
derivative of the corresponding spectra) is also consistent with that
seen in CCOC. We also took a limited amount of data on Sr$_2$CuO$_2$Br$_2$
(not shown).  The general features of Sr$_2$CuO$_2$Br$_2$ were
consistent with that of the other oxyhalides, although significant
charging was observed.  The data on the oxybromides suggests that the low
energy excitations are relatively independent of the apical atoms and the
observed spectral function in ACOC thus originates from the half filled
CuO$_{2}$ plane.  Consequently, the comparison of ARPES data on ACOC at
half filling with Bi2212 data at finite doping seems valid.

\section{B\lowercase{i$_2$}S\lowercase{r$_2$}E\lowercase{r}C\lowercase{u$_2$}O\lowercase{$_{8}$} 
and 
B\lowercase{i$_2$}S\lowercase{r$_2$}D\lowercase{y}C\lowercase{u$_2$}O\lowercase{$_{8}$} 
}

\begin{figure}[bt]
	\centering
	\includegraphics[width=3in]{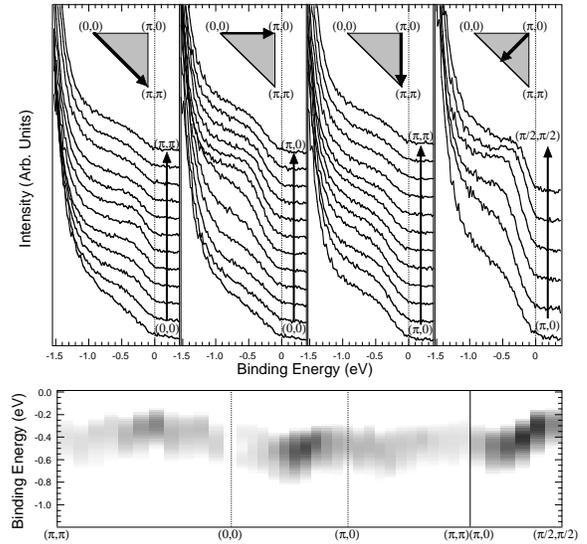}
	\vspace{0.1in}
\caption[ARPES spectra and second derivative plot of 
	Bi$_2$Sr$_2$ErCu$_2$O$_{8}$]{EDCs of Bi$_2$Sr$_2$ErCu$_2$O$_{8}$ 
	along the high symmetry directions as indicated
in the cartoons. Below is a plot of the second derivative
of the above EDCs, from which one can trace out the dispersion.
$E_{\gamma}$=22.4eV and T=100K.}
	\label{BSECOhighSym}
\end{figure}

\begin{figure}[bt]
	\centering
	\includegraphics[width=3in]{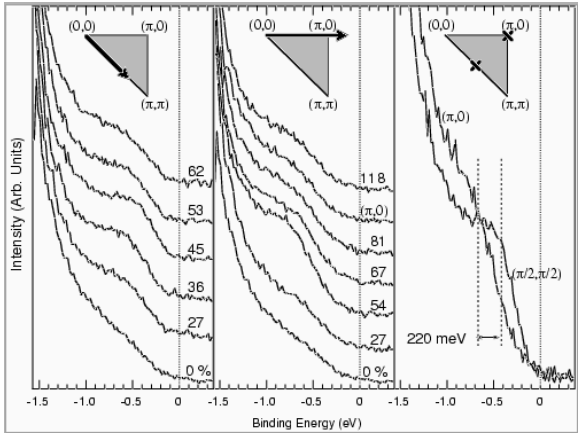}
	\vspace{0.1in}
\caption[ARPES spectra of 
	Bi$_2$Sr$_2$DyCu$_2$O$_{8}$]{EDCs of Bi$_2$Sr$_2$DyCu$_2$O$_{8}$ 
	along the high symmetry directions as indicated
in the cartoons. $E_{\gamma}$=22.4eV and T=75K. The dashes lines in 
the right panel mark the positions of maximum curvature.}
	\label{BSDCOhighSym}
\end{figure}

Of course, as mentioned previously, the ideal scenario to examine the
doping evolution is to study the same system through the entire doping
range.  Therefore, we also present Er and Dy doped Bi2212 crystals grown
near half filling.  Figures \ref{BSECOhighSym} and \ref{BSDCOhighSym} show
the EDCs for these samples along the high symmetry directions.  The low
energy features are not nearly as well defined as in the oxyhalides;
however, a clear shoulder does emerge from the background.  The poor
definition of these features is consistent with previous studies on
underdoped Bi2212 which show that the low energy excitation spectra become
smeared out as one proceeds toward half filling in this
system.\cite{MarshallPRL96}, \cite{ShenPhysRep95} This is precisely the
reason why ARPES data on the oxychlorides has been so valuable to
understanding the problem of a single hole in an antiferromagnet.  The
reason for the relatively poor spectral quality of heavily underdoped
Bi2212 is an open question, although it is possibly simply an issue of
sample quality, such as lattice strain.

Examining the low energy spectra from (0,0) to $(\pi,\pi)$ of Er doped
Bi2212 we see that the shoulder is most pronounced at $(\pi/2,\pi/2)$ with
a minimum binding energy of -0.27 eV. From (0,0) to $(\pi,0)$ the shoulder
develops at higher binding energy with a value of -0.49 eV at $(\pi,0)$
and then disappears again as one travels from $(\pi,0)$ to $(\pi,\pi)$.  
Although there is significant ambiguity in identifying the binding energy
of these features, the fact that the shoulder has a dispersion of roughly
0.22eV between $(\pi,0)$ and $(\pi/2,\pi/2)$ can be clearly seen in the
final panel.  The second derivative of the spectra from which the
above values are attained are also presented in figure
\ref{BSECOhighSym}. One can see that the dispersion found with this
method reproduces the above description.  Results on Dy doped Bi2212 shown
in figure \ref{BSDCOhighSym} clearly mimic the behavior seen in the Er
doped Bi2212 including the 0.22 eV difference in energy between
$(\pi/2,\pi/2)$ and $(\pi,0)$.  The overall features of the dispersion and
the resulting energy difference of roughly 220 meV between $(\pi/2,\pi/2)$
and $(\pi,0)$ in Er and Dy doped Bi2212 is qualitatively consistent with
the dispersion and the 350 meV $d$-wave-like modulation seen in
CCOC. We suspect the quantitative difference between Bi2212 and ACOC
may simply be due to the poor definition of spectral features of the
former, but we can not rule out the possibility that it is an intrinsic
difference between the two samples. Nevertheless, even with the poor
spectral definition in heavily underdoped Bi2212, the electronic
structure is still qualitatively consistent between the oxyhalide and
Bi2212 systems.  This result along with our data on CCOB, justifies the
long standing assumption that the ARPES data from SCOC and CCOC are
representative of a half filled CuO$_{2}$ square lattice.

\section{S\lowercase{r$_2$}C\lowercase{u$_3$}O\lowercase{$_4$}C\lowercase{l$_2$}: 
C\lowercase{u$_{3}$}O\lowercase{$_{4}$} plane}

\begin{figure}[tb]
	\centering
	\includegraphics[width=3in]{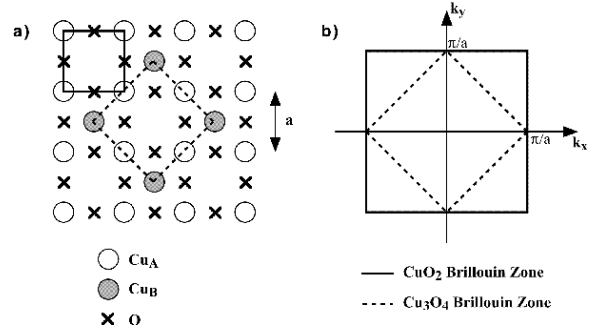}
	\vspace{0.1in}
\caption[Cartoon comparison of CuO$_2$ and Cu$_3$O$_4$ unit 
	cells]{Real space(a) and Reciprocal space(b) cartoons of the unit 
	cell of Cu$_{3}$O$_{4}$(dashed line). The unit cell of the 
	CuO$_{2}$(solid line) is shown for comparison. }
	\label{Cu3O4BZ}
\end{figure}

Sr$_2$Cu$_3$O$_4$Cl$_2$ is a particularly remarkable example of the 
apparent robustness of the structure of the lowest energy excitations.  
This system deviates from the other cuprates due to an additional Cu 
atom located in every other CuO$_{2}$ plaquette resulting in an 
in-plane stoichiometry of Cu$_{3}$O$_{4}$.  The resulting crystal 
structure has a unit cell twice as large as in Sr$_2$CuO$_2$Cl$_2$ and 
rotated by 45 degrees (see figure \ref{Cu3O4BZ}).  Thus $(\pi,0)$ of 
the CuO$_{2}$ unit cell is now equivalent to $(\pi,\pi)$ in the 
Cu$_{3}$O$_{4}$ system, and similarly $(\pi/2,\pi/2)$ for 
Sr$_2$CuO$_2$Cl$_2$ is equivalent to $(\pi,0)$ for 
Sr$_2$Cu$_3$O$_4$Cl$_2$.  However, instead of presenting results in 
the new Cu$_{3}$O$_{4}$ basis, we will continue to present all 
momentum values in the original CuO$_{2}$ basis.  This will make for a 
simpler comparison between the two systems.

\begin{figure}[b!]
	\centering
	\includegraphics[width=3in]{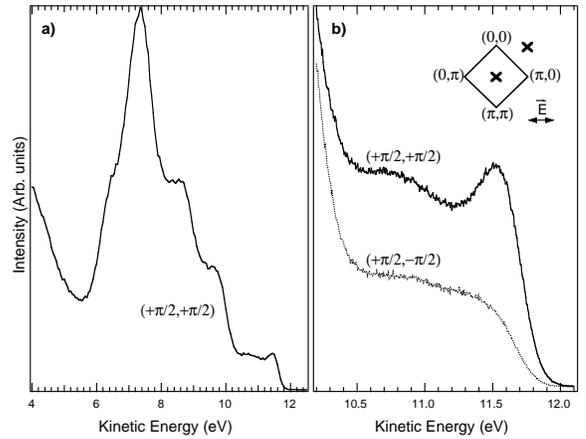} 
	\vspace{0.1in}
\caption[Valence 
	band spectra of Sr$_{2}$Cu$_{3}$O$_{4}$Cl$_{2}$ at high symmetry 
	points]{a) Valence band spectra of Sr$_{2}$Cu$_{3}$O$_{4}$Cl$_{2}$ 
at 
	$(+\pi/2,+\pi/2)$.  b) An expanded view of the low energy spectra.  
	A comparison of the spectra at $(+\pi/2,+\pi/2)$(solid line) and 
	$(+\pi/2,-\pi/2)$(dotted line) indicates that the wavefunction 
	responsible for the low energy excitations is odd with respect to 
	a 45$^{\circ}$ line relative to the Cu-O bond direction.  
	$E_{\gamma}$=16.55eV and T=293K.}
	\label{Sr2342VBpol}
\end{figure}

A sample valence band spectra of Sr$_2$Cu$_3$O$_4$Cl$_2$ at 
$(\pi/2,\pi/2)$ is shown in figure \ref{Sr2342VBpol}.  In fact it is 
very similar to the valence band of Sr$_2$CuO$_2$Cl$_2$ at 
$(\pi/2,\pi/2)$.\cite{DurrPRB00} We will however, continue to focus on 
the ``foot'' of the valence band which contains the low energy 
excitations.  The following panel examines the low energy excitation 
spectra at two equivalent points: $(+\pi/2,+\pi/2)$ and 
$(+\pi/2,-\pi/2)$.  Since the polarization of the incident light is 
horizontal, the suppression of spectral weight at $(+\pi/2,-\pi/2)$ 
relative to $(+\pi/2,+\pi/2)$ indicates that the wave function at 
$(\pi/2,\pi/2)$ is odd with respect to a mirror plane at 45$^{\circ}$ 
to the Cu-O bond direction.  This observation is consistent with the 
state having a $d_{x^{2}-y^{2}}$ orbital character.

\begin{figure}[bt]
	\centering
	\includegraphics[width=3in]{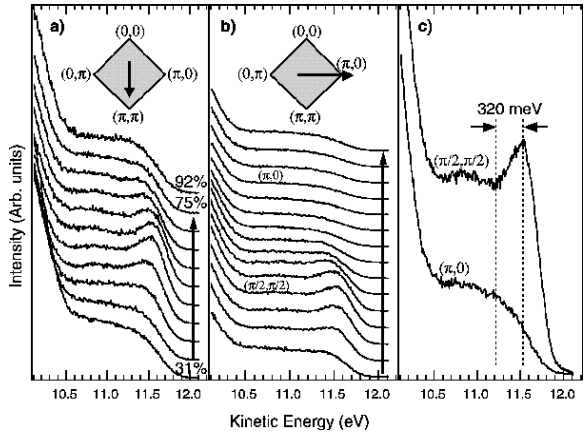} 
\vspace{0.1in} 
	\caption[ARPES spectra of Sr$_2$Cu$_3$O$_4$Cl$_2$]{a) and b) EDCs 
	of Sr$_2$Cu$_3$O$_4$Cl$_2$ along two high symmetry directions as 
	indicated in the inset.  c) Comparison of the spectra at 
	$(\pi/2,\pi/2)$ and $(\pi,0)$.  $E_{\gamma}$=16.5eV and T=293K.}
	\label{Sr2342highSym}
\end{figure}

In figure \ref{Sr2342highSym}, two perpendicular cuts through
$(\pi/2,\pi/2)$ are shown.  Note that the lowest energy feature is most
well defined near $(\pi/2,\pi/2)$, the dispersion is isotropic about its
maximum at $(\pi/2,\pi/2)$, and as can be seen in the final panel,
$(\pi,0)$ lies approximately 320 meV below $(\pi/2,\pi/2)$, although
significant uncertainty exists in the peak position. These
results, including the polarization dependence, are identical to the case
of ACOC.\cite{DurrPRB00} They are also consistent with earlier reports on
the isostructural compound Ba$_2$Cu$_3$O$_4$Cl$_2$ taken at a higher
photon energy.\cite{GoldenPRL97,ShmelzPRB98} We confirm the observation
that the spectral features about $(\pi/2,\pi/2)$ have a remarkable
resemblance to those in Sr$_2$CuO$_2$Cl$_2$. It is astounding that the
properties of the Zhang-Rice singlet seen in ACOC are almost undisturbed
by the drastic change to the CuO$_{2}$ lattice in the case of
Sr$_2$Cu$_3$O$_4$Cl$_2$.

\section{E$_{\gamma}$ Dependence on E(\lowercase{$k$}) and 
\lowercase{n}(\lowercase{$k$})}

Having shown that the low energy spectra of a half-filled CuO$_{2}$ plane
are independent of the material system being studied, we next attempt to
separate the features of the data which are representative of the single
particle spectral function, A(${\bf k}, \omega$), from those which are a
result of the matrix elements.  Unfortunately, extracting the single
particle spectral function from ARPES measurements is difficult due to the
fact that the measured photoemission intensity under the sudden
approximation is a product of the occupied single particle spectral
function and the matrix element.  In interacting electron systems, it is
impossible to calculate the matrix elements exactly, thus further
complicating the ARPES analysis.  Although we note that symmetry arguments
can be very powerful in understanding some properties of the matrix
element,\cite{DurrPRB00} in general it is a function of the experimental
geometry, photon energy, and the electronic wave function.  Since its
details are not well understood, the objective in a given photoemission
study must be to focus on only those features of the data which are robust
against variations in the experimental conditions.

\begin{figure*}[tb]
        \centering
        \includegraphics[width=6in]{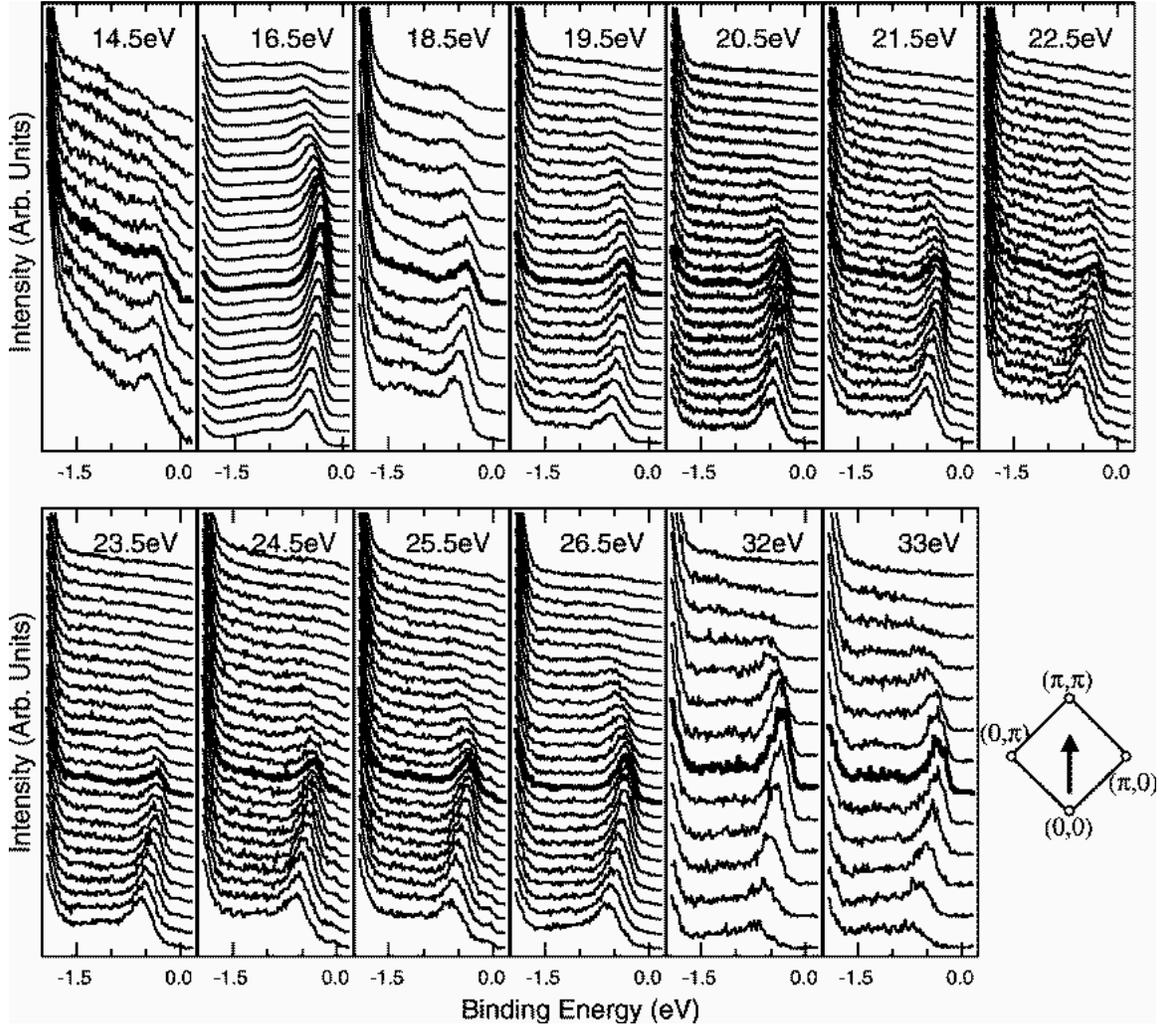}
        \vspace{0.1in}
\caption[Ca$_2$CuO$_2$Cl$_2$ EDCs111b along
        $\Gamma\rightarrow(\pi,\pi)$ using 14.5eV to 33eV photons]{EDCs
from a single cleave of CCOC along $\Gamma\rightarrow(\pi,\pi)$ for 13
different photon energies indicated in each panel respectively. The bold
spectra indicates $(\pi/2,\pi/2)$. The angular separation between the top 
and bottom spectra in each panel is 11.4$^{\circ}$ with an energy
resolution, $\Delta E\leq 50$meV. T=200K.}
        \label{EDCsCaCu111b}
\end{figure*}

Here we present a photon energy dependence study to extract the 
intrinsic E($k$) and n($k$) structure from the raw ARPES data which is 
modulated by the matrix element.  On a single cleave we measured EDCs 
along the nodal direction for 13 different photon energies from 14.5 
to 33eV which are shown in figure \ref{EDCsCaCu111b}.
The morphology of the spectra are similar 
to one another. There are two aspects we wish to address.  
First, the minimum binding energy occurs near $(\pi/2,\pi/2)$ for all 
the photon energies studied, and second, with a few exceptions, the 
intensity profile begins to lose weight before the minimum in binding 
energy is reached (photon energies of 16.5, 32, and 33eV are the 
exceptions where the intensity is symmetric about $(\pi/2,\pi/2)$).  
We will address the latter point in more detail below.  

\begin{figure}[b!]
	\centering
	\includegraphics[width=3in]{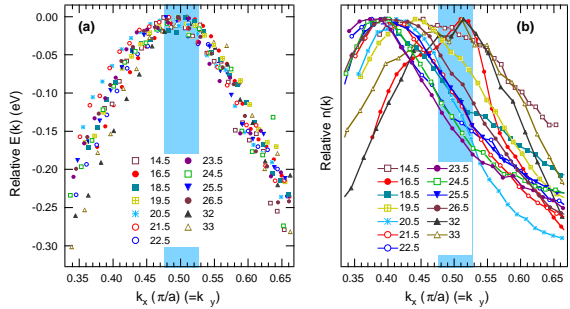}
	\vspace{0.1in}
\caption[Ca$_2$CuO$_2$Cl$_2$ $E({\bf k})$ and $n({\bf k})$ 
	dependence on photon energy]
	{a)$E({\bf k})$ obtained along the nodal direction from the 13
photon energies presented in figure \ref{EDCsCaCu111b}. $E({\bf k})$ is 
	determined by the minimum of the second derivative of the EDCs.  
	b) n(${\bf k}$) obtained from the same data by integrating the 
	EDCs over an 800meV window.  The maxima of each curve were 
	normalized to each other for display.  The shaded bar
represents the momenta error 
	bars.}
	\label{EkEphotsum}
\end{figure}

To check if the dispersion is indeed independent of photon energy we plot
the peak position versus $k$ for each photon energy in figure
\ref{EkEphotsum}a.  The peak positions were found by taking the minimum of
the second derivative of each spectra.  One can see that to within our
experimental limits, which were determined by the reproducibility of the
dispersion on subsequent scans under identical conditions and is given by
the width of the blue bar, the dispersion is independent of photon energy.  
This agrees with most of the previous reports on
SCOC\cite{WellsPRL95,KimPRL98,PothuizenPRL97}, including one very
detailed, recent study\cite{DurrPRB00}, but contrast with the results from
Refs.  \cite{HaffnerPRB01,LaRosaPRB97}, the former of which report that
the minimum binding energy position shifts by approximately 10$\%$ of the
(0,0) to $(\pi,\pi)$ distance to $(0.39\pi,0.39\pi)$ when using 35eV
photons.  (note that this is outside of our momenta error bars) They
attribute this change to matrix elements where the dependence on binding
energy varies as a function of photon energy.

\begin{figure*}[t!]
	\centering
	\includegraphics[width=5.5in]{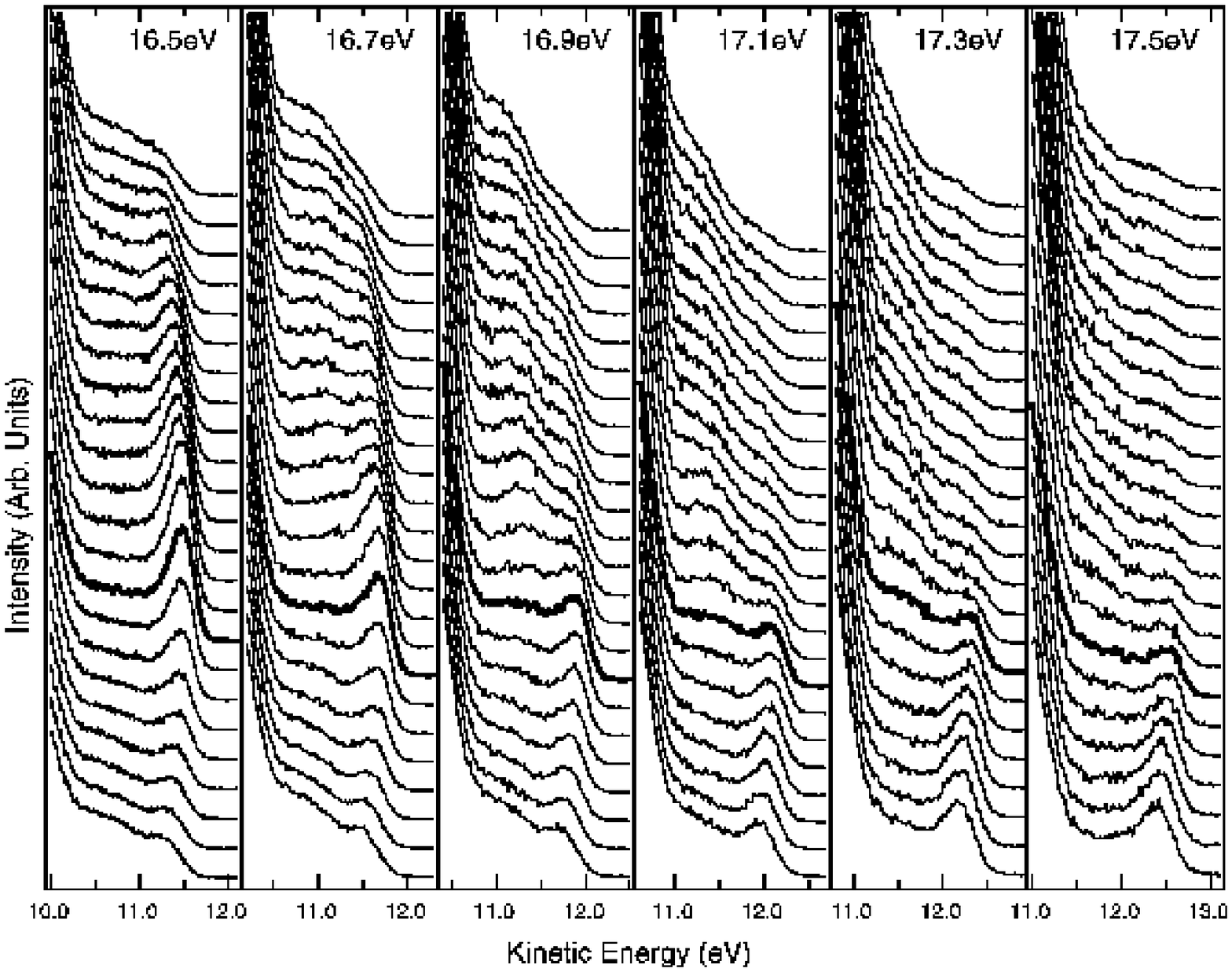}
	\vspace{0.1in}
\caption[Ca$_2$CuO$_2$Cl$_2$ EDCs along 
	$\Gamma\rightarrow(\pi,\pi)$ using 16.5eV to 17.5eV photons]{EDCs from a single cleave of CCOC along the nodal 
	direction for $E_{\gamma}$=16.5eV to $E_{\gamma}$=17.5eV as indicated 
	in each panel. The bold spectra indicates $(\pi/2,\pi/2)$. The angular 
	separation between the top and bottom spectra in each panel is 
	13$^{\circ}$  with an energy resolution, $\Delta E=70$meV. T=293K.}
	\label{Ephot2comp}
\end{figure*}

Aside from indicating the expected two dimensional nature of the
dispersion, our photon energy dependence has also clearly resolved the
presence of a second component in the low energy electronic structure of
the half filled insulator.\cite{KimPRL98} Figure \ref{Ephot2comp} presents
EDCs in the nodal direction for six photon energies from 16.5eV to 17.5eV.
Aside from the feature typically associated with the Zhang-Rice singlet, a
second component is observed at approximately 600meV higher binding
energy.  In the nodal direction this feature is most clearly resolved at
17eV. From this data it is clear that when attempting to model the data on
the insulator, one can not simply treat the high energy spectral weight as
a featureless background.  Although the dispersion of this feature is
difficult to track it appears to mimic the dispersion of the lowest energy
feature.  The strong influence of matrix elements and a lack of
understanding of the origin of the spectral lineshapes prevents us from
extracting detailed information on this high energy feature and its
evolution with photon energy. One possible origin of this feature is from
string resonances.\cite{DagottoRMP94} String resonances occur when a hole,
created in an antiferromagnetic background, experiences a confining
potential due to the energy cost associated with disrupting the
antiferromagnetic order as it hops away from its original location.  The
lowest energy state for the hole in this potential corresponds to the
Zhang-Rice singlet, while the first excited state is predicted to lie
roughly 1.8$(J/t)^{2/3}\approx0.5$eV higher in
energy.\cite{StringResonance} Although we are not certain of the origin of
the higher energy feature, the string resonance concept provides an
intriguing possibility for further study.

\begin{figure}[btp]
	\centering
	\includegraphics[width=3in]{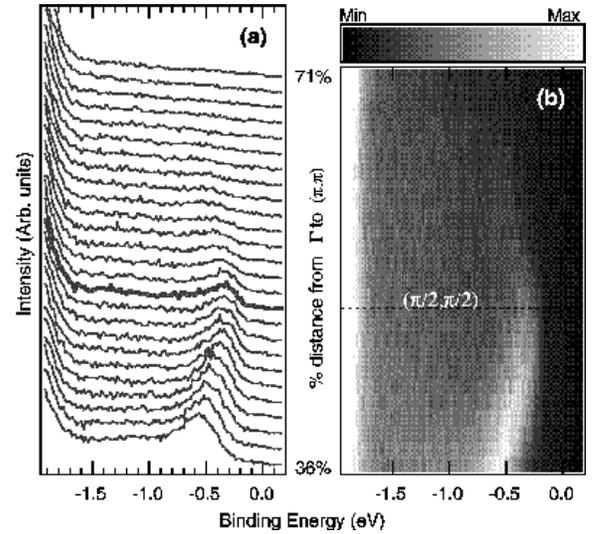} 
	\vspace{0.1in}
\caption[An example of an asymmetric spectral intensity about 
	$(\pi/2,\pi/2)$ in Ca$_2$CuO$_2$Cl$_2$]{a) EDCs and 
	b) corresponding image 
	plot of CCOC taken along the nodal direction with 23.5eV photons.  
	The bold spectra indicates $(\pi/2,\pi/2)$, and T=200K.}
	\label{NodalCut23eV}
\end{figure}

Having examined the dispersion, we now look at the more difficult problem
of extracting the underlying n(${\bf k}$) structure.  It is of particular
interest to determine whether or not the underlying intensity profile of
A(${\bf k}, \omega$) has an asymmetry with respect to the
antiferromagnetic zone boundary as this would put constraints on the valid
coupling regime for different models.\cite{DagottoRMP94,EskesPRB96} For
the majority of data shown in figure \ref{EDCsCaCu111b} we noticed that
the intensity of the lowest energy excitation begins to lose weight before
$(\pi/2,\pi/2)$. By examining the image plot of a representative set of
spectra (as done in figure \ref{NodalCut23eV}) this effect can be seen to
be quite dramatic. This has also been observed previously by several
authors with isolated photon
energies.\cite{HaffnerPRB00,WellsPRL95,KimPRL98,PothuizenPRL97,LaRosaPRB97}
In figure \ref{EkEphotsum}b the n(${\bf k}$) curves obtained by
integrating the EDCs from -0.5 to 0.3eV relative to the valence band
maximum for each photon energy are shown. In presenting n(${\bf k}$) we
implicitly use the sudden approximation to extract the momentum
distribution function n(${\bf k}$), from ARPES data via the relation
n(${\bf k}$) = $\int A(\vec{k},\omega)f(\omega)  d\omega$ where
$f(\omega)$ is the Fermi function.\cite{RanderiaPRL95} We note here that
the experimental quantity measured is not precisely n(${\bf k}$), as the
photoemission intensity is weighted by the matrix element and the
integration window is limited due to the presence of additional bands.  
The energy integration window was chosen so as to minimize the
contribution from the second component seen at higher binding energy, but
we note that the results are qualitatively independent of the specific
energy window chosen.

Clearly, with the exception of 16.5, 32, and 33eV the intensity profile
peaks well before $(\pi/2,\pi/2)$, and drops rapidly as one crosses the
antiferromagnetic zone boundary at $(\pi/2,\pi/2)$.  The fact that 16.5,
32 and 33eV data can differ so dramatically is evidence of what the matrix
elements can do.  If the underlying spectral function were symmetric about
$(\pi/2,\pi/2)$ then one would expect that the distribution of curves in
figure \ref{EkEphotsum}b would be evenly distributed about
$(\pi/2,\pi/2)$.  This is the case for a similar study on SCOC, which used
five photon energies from 20 to 24eV.\cite{HaffnerPRB01} They observe the
intensity profile to be more heavily weighted towards (0,0) at
$E_{\gamma}$=24eV which then gradually shifts until at $E_{\gamma}$=20eV
the profile is more heavily weighted towards $(\pi,\pi)$.  From this they
conclude that the true underlying n(${\bf k}$) is symmetric.  However, the
data in figure \ref{EkEphotsum} from 13 different photon energies suggests
otherwise.  10 curves have an asymmetry more heavily weighted towards
(0,0), 3 are peaked at $(\pi/2,\pi/2)$, and none are more heavily weighted
towards $(\pi,\pi)$.  Combining figure \ref{EkEphotsum}b with the majority
of other single photon energy studies on
SCOC\cite{HaffnerPRB00,WellsPRL95,KimPRL98,PothuizenPRL97,LaRosaPRB97}
suggests to us that a true asymmetry does exist in the underlying
occupation probability, n(${\bf k}$).

\begin{figure}[t!]
	\centering
\includegraphics[width=3in]{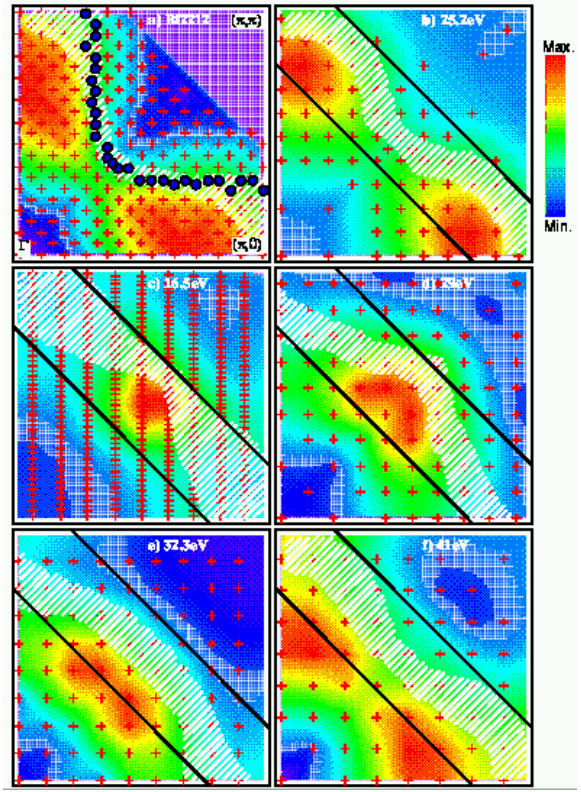} 
\vspace{0.1in}
\caption[$E_{\gamma}$ dependence of n(${\bf k}$) over the entire 
Brillouin zone]{(color) 
Integrated spectral weight.  The crosses indicate where spectra were 
taken.  Except for (c) the data is symmetrized about the 
$k_{x}$=$k_{y}$ line.  Red is maximum.  (a) optimally doped Bi2212 at 
$E_{\gamma}$=22.4eV. The white hashed region indicates the approximate 
location of the Fermi surface determined from n(${\bf k}$).  The dots 
illustrate the position of the Fermi surface as determined by the 
traditional method for analyzing ARPES data.  (b) CCOC shows a 
striking similarity to the metal allowing the 
identification of the white hashed region as a remnant Fermi surface.  
Comparison of (b) through (f) show CCOC taken at photon energies of 
25.2, 16.5, 29, 32.3, and 41eV. The intensity maxima varies between 
different panels, but the loss of intensity as one approximately 
crosses the antiferromagnetic zone boundary is a consistent feature.  
The cumulative boundary of the remnant Fermi surface is drawn with 
black lines on panels (b) through (f). Resolution $\Delta E\leq 
70$meV.}
	\label{Ephotnkmaps}
\end{figure}

As all of the above studies were only from a single cut through the
Brillouin zone we now examine a more global perspective of the intensity
profile.  Figure \ref{Ephotnkmaps}b)-f), presents relative n(${\bf k}$)
patterns of CCOC over a Brillouin zone quadrant for five different photon
energies while panel a) is of an optimally doped Bi2212 sample.  Spectra
are taken at the crosses, values in between were generated from a linear
interpolation, and except for panel (c), the data has been symmeterized
about $k_{x}$=$k_{y}$ and the geometry of the experimental setup was
identical for each data set.  The in-plane component of the electric field
was polarized along the Cu-O bond direction (This is a 45$^{\circ}$
rotation relative to the data presented in figures \ref{EDCsCaCu111b} and
\ref{Ephot2comp}). The first two panels reproduce the original comparison
of metallic Bi2212 and insulating CCOC from which the initial
identification of a ``remnant Fermi surface'' was
made.\cite{RonningScience98} For a state with $d_{x^{2}-y^{2}}$ orbital
symmetry the suppression of weight as one approaches the line $k_{x}$=$0$
is expected.  In Bi2212 the only drop in intensity which is not naturally
explained by matrix elements is where a Fermi surface crossing has
occurred.  As seen in the figure the intensity drop matches the
traditional method for determining a Fermi surface crossing by following
the dispersion by eye and is indicated by the dots.  This indicates that
the surface of steepest descent in the experimentally determined n(${\bf
k}$) is a valid method of determining the Fermi surface.  The surprise is
that a similar drop in intensity is observed in the insulator.  Although
the feature is less well defined here, the striking resemblance it bears
to the metal suggests that the origin is similar, and hence it was
qualitatively described as a remnant Fermi surface.\cite{RonningScience98}
[Note that the remnant Fermi surface is still fully gapped by the large
Mott gap.]

Here we examine the effect of changing the photon energy to determine if
this feature is robust.  From panels (b) through (f) one immediately
notices that the intensity pattern varies tremendously for the five photon
energies: 25.2, 16.5, 29, 32.3, and 41eV. However, the variations appear
predominantly parallel to the $(\pi,0)$ to $(0,\pi)$ direction, while
perpendicular to this there exists relatively little variation as we noted
above in figure \ref{EkEphotsum}.  The exact shape of the remnant Fermi
surface may change, but at all photon energies used there is a loss of
spectral weight as one crosses the approximate antiferromagnetic zone
boundary from $(0,\pi)$ to $(\pi,0)$.  It may appear that the remnant
Fermi surface crosses the (0,0) to $(\pi,0)$ cut or the $(\pi,0)$ to
$(\pi,\pi)$ cut depending on the photon energy chosen, but the broadness
and the variability due to matrix elements prevent one from clearly
identifying the intensity profile as either case.  However, the data
clearly suggest that globally, there is an asymmetry in intensity stronger
towards (0,0) than towards $(\pi,\pi)$ as one crosses the region spanned
by the black lines in the Brillouin zone which is coincident with the
antiferromagnetic zone boundary.

\begin{figure}[tb]
	\centering
\includegraphics[width=3in]{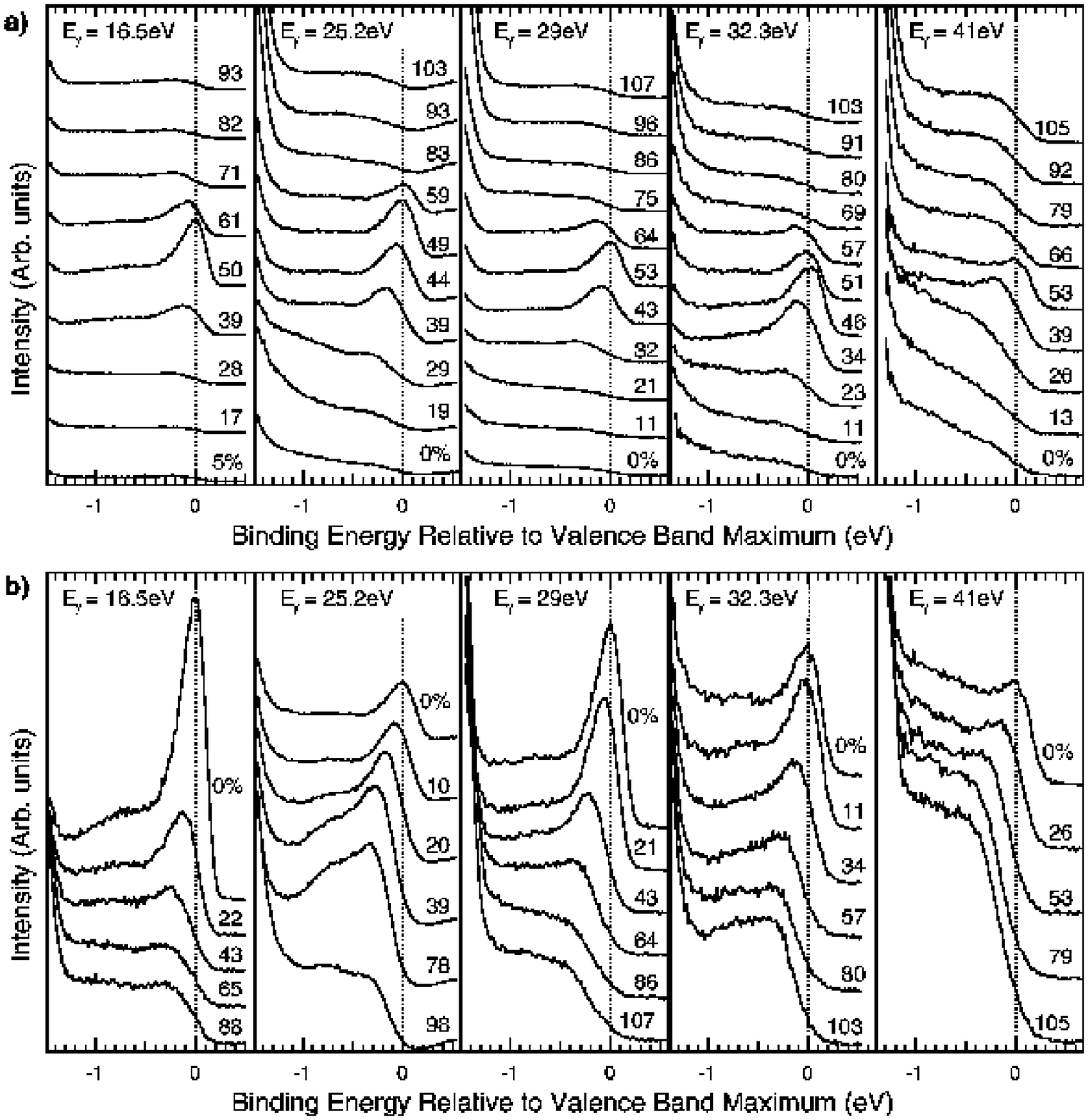} 
\vspace{0.1in} 
\caption[ARPES spectra from the n(${\bf k}$) mappings of 
Ca$_2$CuO$_2$Cl$_2$]{ARPES spectra along 2 cuts, a) (0,0)(bottom) to 
$(\pi,\pi)$(top) and b) $(\pi/2,\pi/2)$(top) to $(\pi,0)$(bottom), with 5 
different photon 
energies (16.5, 25.2, 29, 32.3, and 41eV). The numbers refer to the 
percentage distance along each cut. The intensities of the 
features vary, but the dispersion remains the same.}
	\label{EphotEDCBL53}
\end{figure}

\begin{figure}[tb]
	\centering
\includegraphics[width=3in]{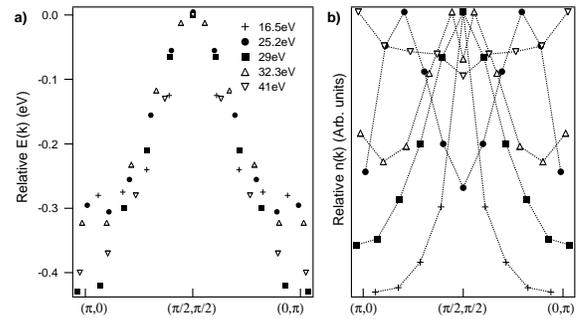} 
\vspace{0.1in} 
\caption[Comparison of $E({\bf k})$ and n(${\bf k}$) dependence on 
$E_{\gamma}$ $\parallel$ to the antiferromagnetic zone 
boundary]{(a) Peak position and (b) n(${\bf k}$) from the 
spectra in figure \ref{EphotEDCBL53} along the antiferromagnetic zone 
boundary.  The data along $(\pi,0)$ to $(0,\pi)$ has been symmetrized.  
Again the dispersion is independent of photon energy while the 
intensity varies randomly indicating that this modulation in intensity 
is solely due to matrix elements.}
	\label{EphotsumBL53}
\end{figure}

Although the n(${\bf k}$) image plots can provide a wealth of information
and are extremely good for summarizing data, it is important to also look
at the raw data to fully appreciate the information being given by the
image plots.  This is done in Figure \ref{EphotEDCBL53}.  Panels (a) and
(b) plot respectively, the EDCs from (0,0) to $(\pi,\pi)$ and
$(\pi/2,\pi/2)$ to $(\pi,0)$, from the data sets used to create the
intensity maps in figure \ref{Ephotnkmaps}b)-f).  We find the spectra are
qualitatively similar.  This is true even at 41eV where the peak is poorly
defined throughout the zone.  To examine them more closely, Figure
\ref{EphotsumBL53} plots both the dispersion of the peak position and
n(${\bf k}$) together for all the cuts along the antiferromagnetic zone
boundary.  Similar to the nodal direction shown in figure
\ref{EkEphotsum}a, only slight differences exist in dispersion, $E({\bf
k})$, among the five different photon energies.  However, the spectral
intensity, n(${\bf k}$), varies seemingly randomly with photon energy.  
In the extreme case between 25.2eV and 29eV the intensity is increasing as
one approaches $(\pi/2,\pi/2)$ for the former, and decreasing for the
latter.  One might expect such behavior when the underlying spectral
function does not possess an asymmetry.  Meanwhile the asymmetry in
spectral weight perpendicular to the antiferromagnetic zone boundary (see
figure \ref{EkEphotsum}b) appears robust to the strong variations in the
matrix element.

In general, the outgoing photoelectrons resulting from photons with
different wavelengths will necessarily have a different overlap with the
initial wavefunctions of the system, and hence different cross-sections.  
This is impossible to avoid.  Ideally, to eliminate such matrix element
effects one would average over all possible photon energies, and
experimental geometries.  Here we have observed large variations along the
antiferromagnetic zone boundary that are clear manifestations of the
matrix element, while the asymmetric shift of weight towards (0,0) with
respect to the $(\pi,0)$ to $(0,\pi)$ line suggests that this is a
property of the underlying spectral function.  Our initial report implied
that the underlying n(${\bf k}$) structure matched that of the LDA Fermi
surface at half filling without correlations.\cite{RonningScience98} It is
now clear, that the matrix elements are strong enough to make such a
precise identification very difficult.  However, we verify that an
underlying asymmetry exists in the spectral function about the
antiferromagnetic zone boundary which is robust despite variations with
photon energy.  Whether it truly lies along the LDA Fermi surface, the
antiferromagnetic zone boundary, or some other contour is a more difficult
question to answer.

\section{Rounded Nodal Dispersion}

The greatest utility of the remnant Fermi surface concept as it pertains
to the high temperature superconductors is that a $d$-wave-like dispersion
was identified in the insulator, which could thus provide a natural
connection to the $d$-wave-like high energy
pseudogap.\cite{RonningScience98} In ARPES the pseudogap about
$(\pi,0)$ is characterized in two ways. The first is by the use of the
leading edge midpoint, which has the same energy scale as the
superconducting gap, and is typically defined in the same way, only above
T$_c$. However, another energy scale exists in the spectra which is an
order of magnitude larger, and is sometimes refered to as a hump or a high
energy pseudogap. This is the feature which provides a direct connection
with the dispersion observed in the insulator.\cite{pseudogaps} The
remnant Fermi surface identifies a topological contour in the Brillouin
zone of the insulator, which in this case is the antiferromagnetic zone
boundary.  By examining the dispersion along this contour one can identify
a ``gap'' in this system, the dispersion of which along the remnant Fermi
surface fits the d-wave functional form remarkably well.  We now extend
our earlier study on CCOC to examine the precise nature of the dispersion
near the nodal line $k_{x}$=$k_{y}$.\cite{RonningScience98} A simple
$d$-wave dispersion proportional to $|\cos k_{x}a-\cos k_{y}a|$ would
produce a linear dispersion with a discontinuous derivative perpendicular
to the nodal direction at $k_{x}$=$k_{y}$.  This is most easily seen along
the antiferromagnetic zone boundary where the above function reduces to
$|\sin(k_{x}a-\pi/2)|$.  Such a non-analytic dispersion is non-trivial,
and hence the presence or absence of such a dispersion is of great
significance to theories which attempt to unify the antiferromagnetic
insulator with the $d$-wave superconductor.

\begin{figure}[tbp]
	\centering
	\includegraphics[width=3in]{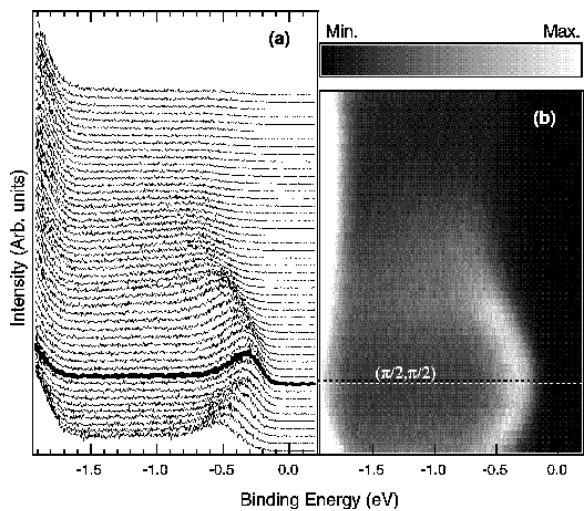}
	\vspace{0.1in}
\caption[ARPES spectra of Ca$_2$CuO$_2$Cl$_2$ along $(\pi,0)$ to 
	$(0,\pi)$]{(a) EDCs and (b) corresponding image plot of 
Ca$_2$CuO$_2$Cl$_2$ 
	data taken along the antiferromagnetic zone boundary at 
	0.6$^{\circ}$ intervals. $(\pi/2,\pi/2)$ is indicated by the bold EDC 
	and the dotted line in the image plot. Data collected at T=200K, 
25.5eV photons, Resolution $\Delta E=15$meV, and 0.8$^{\circ}$ 
angular resolution}
	\label{RoundNodeEDCs}
\end{figure}

\begin{figure}[tb]
	\centering
	\includegraphics[width=3in]{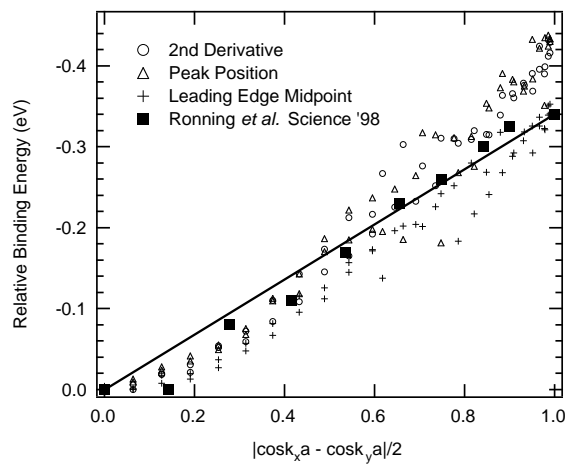}
	\vspace{0.1in}
\caption[$d$-wave comparison of the detailed $E({\bf k})$ of Ca$_2$CuO$_2$Cl$_2$]
	{Examining the curvature about the nodal line using three 
methods for characterizing the dispersion of the data in 
	figure \ref{RoundNodeEDCs} plotted with the original data from 
	ref.  \cite{RonningScience98}.  A straight line corresponds to a 
	simple $d$-wave dispersion.  Each curve was set to zero relative 
binding energy at the 
	node.  Note that the $d$-wave-like gap is in addition to the Mott 
	gap which is not shown.  The increased scatter as one approaches 
	$(\pi,0)$ is indicative of the increased difficulty in tracing the 
	dispersion away from $(\pi/2,\pi/2)$.}
	\label{Rounddwave}
\end{figure}

In figure \ref{RoundNodeEDCs} we present EDCs taken at 0.6$^{\circ}$
intervals along the antiferromagnetic zone boundary and through
$(\pi/2,\pi/2)$.  One observes a smooth round dispersion through
$(\pi/2,\pi/2)$.  This is even more evident in the image plot of the same
data.  To compare with the $d$-wave functional form we must quantify the
dispersion seen in the raw spectra.  The low energy features seen in the
insulator are inherently very broad with a full width at half maximum at
$(\pi/2,\pi/2)$ of 300meV, which is roughly the total dispersion seen in
this material.  Due to this, and the fact that the higher energy spectral
weight is also dispersive and of unknown origin, the significance of any
particular fit to the data is questionable.  For this reason we have
chosen to quantify the dispersion using three methods: the peak maximum,
the location of maximum curvature, and the leading edge midpoint.  We
compare these quantities against $|\cos k_{x}a-\cos k_{y}a|$ in figure
\ref{Rounddwave}.  The straight line represents the simple $d$-wave
scenario.  Near the node it is clear that the dispersion is rounded in
each case.

\begin{figure}[b!]
	\centering
	\includegraphics[width=3in]{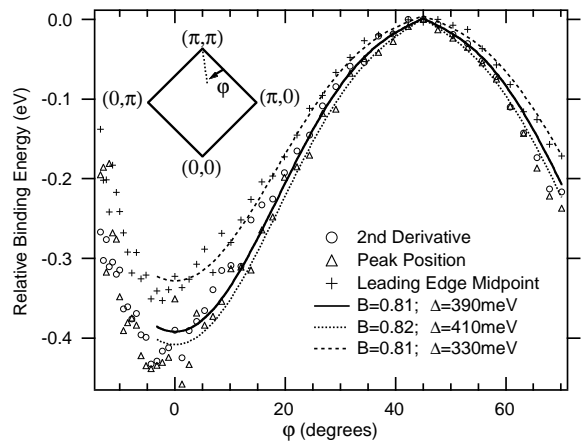} 
	\vspace{0.1in}
\caption[Parameterizing the flattened dispersion near $(\pi/2,\pi/2)$]
	{Fits of the data presented in 
	figure \ref{Rounddwave} to $\Delta|B\cos(2\phi) + 
	(1-B)\cos(6\phi)|$ where $\phi$ is indicated in the figure.  $B$ 
	characterizes the deviation from a simple $d$-wave.}
	\label{roundedBfits}
\end{figure}

One plausible explanation for the flatness of the dispersion near
$(\pi/2,\pi/2)$ could be the dirty $d$-wave scenario.  Previous ARPES work
on underdoped Bi2212 samples also found a flattened dispersion near the
node of the superconducting and normal state low energy pseudo
gaps.\cite{LoeserScience96,HarrisPRB96,MesotPRL99} Earlier works suggested
this was consistent with a dirty $d$-wave scenario since a finite density
of states with zero excitation energy would result in a flattening of the
node region.  However, Mesot {\it et al.} argued that flattening due to
impurities can be ruled out since pair breaking due to impurities should
cause the overall magnitude of the gap to decrease, while they observed
the opposite.\cite{MesotPRL99} Furthermore, we note that ACOC is very
stable and most likely very chemically pure as evidenced by many
unsuccessful attempts by many groups to dope the oxychlorides under
atmospheric conditions.\cite{KohsakaThesis} The broadness of the peaks can
not be used as evidence for impurities since the same feature is almost
equally broad in Bi2212 samples which contain nearly resolution limited
quasiparticle peaks in their spectra in the superconducting
state.\cite{KaminskiPRL00} Although one can not rule out the possibility
that the observed dispersion is due to impurities we will, for the
remainder of this paper, assume that the observed rounding is intrinsic to
the undoped insulator, and discuss the consequences.  We defer a full
discussion on the lineshape of the insulator to another
paper.\cite{KimPRB02}

In an attempt to quantify the deviation from a simple $d$-wave picture we
fit the data to $\Delta|B\cos(2\phi) + (1-B)\cos(6\phi)|$ in figure
\ref{roundedBfits}, where $\phi$ is defined in the figure, $\cos(2\phi)$
is approximately $|\cos k_{x}a-\cos k_{y}a|$, and $\cos(6\phi)$ is the
next allowable harmonic for $d$-wave symmetry.  An increase in the
$\cos(6\phi)$ weight (smaller $B$) corresponds to a flatter nodal region.  
Experimentally, we find $B$=0.81$\pm$ 0.01.  This is smaller than any
underdoped sample measured by Mesot {\it et al.}, and thus is consistent
with their interpretation that the rounding is a result of increasing
antiferromagnetic correlations.\cite{MesotPRL99}

Finally we note that the above data was taken at $E_{\gamma}$=25.5eV.  
Identical results were obtained at $E_{\gamma}$=29eV, on a second cleave
of CCOC at $E_{\gamma}$=25.5eV, and on SCOC using 22.4eV photons.  
Improving the angular resolution by a factor of 4 by narrowing the slits
of our analyzer produced the same results.

\section{Discussion}

\subsection{System Independence}

By examining A$_2$CuO$_2$X$_2$, (A=Sr, Ca; X=Cl, Br) and
Bi$_2$Sr$_2$MCu$_2$O$_{8}$ (M=Er, Dy) we found that the dispersion of the
lowest lying excitations in parent cuprates are independent of the
particular system studied.  Even the results of the Cu$_{3}$O$_{4}$
system, Sr$_2$Cu$_3$O$_4$Cl$_2$, were similar to the half-filled CuO$_{2}$
systems.  We further note that, Nd$_{2-x}$Ce$_{x}$CuO$_{4}$ and
La$_{2-x}$Sr$_{x}$CuO$_{4}$ near half filling have been studied as
well.\cite{InoPRB00,ArmitageNCO,Yoshida01} For Nd$_{2}$CuO$_{4}$ a
shoulder on the edge of the valence band exists and exhibits a minimum
binding energy at $(\pi/2,\pi/2)$.\cite{ArmitageNCO} Perhaps this is
the strongest evidence supporting the independence of the low energy
electronic structure of half-filled cuprates on the apical site as it is
completely unoccupied in Nd$_{2}$CuO$_{4}$. In La$_{2}$CuO$_{4}$, a broad
shoulder exists at $(\pi/2,\pi/2)$ at 0.5eV which disperse by roughly
200meV to higher binding energy as it approaches
$(\pi,0)$.\cite{InoPRB00,Yoshida01} It is also interesting to note
that the exchange coupling J, which enters into the parameterization of
the low energy electronic structure, are nearly equivalent in
La$_{2}$CuO$_{4}$(135 meV) and Sr$_2$CuO$_2$Cl$_2$ (125 meV) as determined
from inelastic neutron and two-magnon Raman
scattering.\cite{KastnerRMP98} All of these observations further support
the universality to all cuprates of the dispersion of the lowest energy
feature seen in Sr$_2$CuO$_2$Cl$_2$ and Ca$_2$CuO$_2$Cl$_2$.

The role of crystallographic structure on the low energy electronic
structure and superconductivity is still not well established. It was
believed that the apical oxygen or structural distortion of the CuO$_2$
planes was necessary for superconductivity, but the observation of
superconductivity in Na-doped Ca$_2$CuO$_2$Cl$_2$ and
Sr$_2$CuO$_2$F$_{2+\delta}$ has since rejected these
beliefs.\cite{HiroiNature94,AlMamouriNature94} In Na-doped
Ca$_2$CuO$_2$Cl$_2$ no structural distortion was found down to 10K while
superconductivity was observed as high as 26K. Perhaps the apical Cl
stabilizes a structure with an increased Cu-O-Cu distance relative to
systems with an apical oxygen. Clearly this is not significant enough to
eliminate superconductivity in CCOC with an a-axis lattice constant of
3.87\AA\, but may be responsible for the fact that SCOC, with a=3.97\AA\,
has not yet been successfully doped into a superconductor.\cite{Grande75}
Although band structure claims that the apical halide ion does not
appreciably enter the low energy density of states of superconducting
oxyhalides,\cite{NovikovPRB95} one should note that hole-doped
superconductivity is observed in compounds containing an apical anion,
while electron-doped superconductivity is found in systems missing an
apical atom. This suggests that the role of the apical atom is not
completely understood.

\subsection{E$_{\gamma}$ Dependence}

Returning to the photon energy dependence of the cuprates, let us first
consider the independence of the observed dispersion $E({\bf k})$ on the
choice of photon energy. In principle, there are several reasons which
could cause $E({\bf k})$ to depend on photon energy which we list below.
We know the cuprates are generally believed to be two dimensional
electronic systems.  In reality the wavefunctions have some finite
$\vec{z}$ extent, and even if they are highly localized will have some
finite overlap with neighboring planes.  In fact such a coupling must be
present to create the observed three dimensional long range magnetic order
seen at half filling.  Certainly, if the wave function overlap were large
enough to create a small dispersion as a function of k$_{z}$ a change in
dispersion would be observed with changing photon energy as this varies
k$_{z}$.  Alternatively, if the matrix element had a significant
dependence in the binding energy of the feature which varied with photon
energy, a photon energy dependent dispersion would also result.  Along
similar reasoning, if there are multiple excitation branches there is no
reason to expect that the matrix elements dependence on photon energy
would be the same for each branch.  Most likely an observed variation with
photon energy would be caused by a combination of several of these
factors.  However, in figures \ref{EkEphotsum} and \ref{EphotsumBL53} we
have shown that the dispersion of the insulator is independent of photon
energy to within our experimental uncertainty.  Three dimensionality may
still play a role in causing a small shift in dispersion with photon
energy, but not surprisingly, it is a safe assumption to treat the
electronic structure of this half filled insulator as essentially two
dimensional.

What then is the significance of the observed underlying intensity
modulation along the nodal direction which appears robust to variations in
photon energy?  In the context of specific many-body models such as the
Hubbard model, there is numerical evidence that a structure in n(${\bf 
k}$) survives
even when the on site Coulomb $U$ drives the system insulating, even
though the discontinuity in n(${\bf k}$) which existed in the metal has
been washed out.\cite{DagottoRMP94,EskesPRB96,Bulut} This effect is linked
to the fact that n(${\bf k}$) reflects the underlying Fermi statistics of
the electronic system, where for the specific case of a two dimensional
square lattice that resembles the CuO$_{2}$ planes of the cuprates, there
is a drop in n(${\bf k}$) across a line that is close to the
antiferromagnetic zone boundary.\cite{DagottoRMP94,EskesPRB96} This
differs in the generalized $t$-$J$ models where the structure in
n(${\bf k}$) is washed out.\cite{EskesPRB96,LemaPRB97}

In the metallic state of optimally doped Bi2212 the steepest descent of
n(${\bf k}$) gives a Fermi surface consistent with traditional ARPES
analysis methods despite the complication of matrix element effects as
shown in figure \ref{Ephotnkmaps}.  In our earlier
paper,\cite{RonningScience98} the n(${\bf k}$) pattern of the insulator
CCOC was found to be strikingly similar to the n(${\bf k}$)  pattern seen
in Bi2212.  This realization, coupled with many-body theoretical
calculations on various forms of the Hubbard
model\cite{DagottoRMP94,EskesPRB96}, suggests that the insulator pattern
contains information that is related to a Fermi surface which has been
destroyed by strong electron-electron interactions thus giving a
qualitative concept of a remnant Fermi surface as the surface of steepest
descent in n(${\bf k}$).  In this paper we have found that the remnant
Fermi surface acts to emphasize a robust feature which we observe in the
insulator, although its precise shape is uncertain due to the effects of
the matrix elements.  While this may not be a rigorous definition, as the
Fermi surface is only defined for a metal, this useful concept allows a
practical connection from the pseudogap seen in underdoped cuprates to the
properties of the insulator.\cite{LaughlinPRL97}

Unfortunately, the fact that the antiferromagnetic Brillouin zone boundary
is quite similar to the underlying Fermi surface means that the roles of
magnetism and that of the Fermi statistics of the underlying band
structure in this spectral weight pattern will be uncertain.  Therefore we
suggest here a few ideal tests on the origin of the n(${\bf k}$)  
pattern. One idea is to study a system whose non-interacting Fermi surface
is radically different from the magnetic Brillouin zone boundary.  Since
such a system compatible with ARPES may be difficult to find, perhaps a
more realistic approach will be to study a system similar to the cuprates,
but where the magnetic correlations are much smaller, such that at the
measured temperature, effects due to magnetic correlations could be ruled
out.

We have shown that the loss in intensity as one crosses the
antiferromagnetic zone boundary is a robust feature of the insulator
Ca$_2$CuO$_2$Cl$_2$ which can not be explained solely by matrix
element effects.  However, the photon energy dependence
underscores the qualitative rather than quantitative nature of the
remnant Fermi surface concept.  We argue that much physics can be
learned in spite of the adverse effects which matrix elements can have in
ARPES, as long as care is taken to properly sort out the intrinsic
versus the extrinsic physics.  In particular, the resulting connection
between the $d$-wave like dispersion and the pseudogap in the underdoped
regime is robust. 

\subsection{{\it d}-wave-like Dispersion}

Let us now address the significance of the rounded dispersion at 
$(\pi/2,\pi/2)$.  Numerical calculations show that the $t$-$J$ model 
describes the dispersion from (0,0) to $(\pi,\pi)$ well, but 
incorrectly predicts the energy at $(\pi/2,\pi/2)$ and $(\pi,0)$ to be 
equal \cite{DagottoRMP94}.  The addition of $t^{\prime}$ and 
$t^{\prime\prime}$ corrects this problem\cite{TohyamaReview} and 
for realistic values of these parameters, also gives an isotropic 
dispersion which is scaled by a single parameter 
$J$.\cite{TohyamaJPSJ00} The effect of adding $t^{\prime}$ and 
$t^{\prime\prime}$ to calculations is to destabilize the single hole 
N\'{e}el state about $(\pi,0)$.  The resulting state appears to be a 
spin liquid which may display spin charge separation, but this 
only occurs about $(\pi,0)$.\cite{TohyamaJPSJ00,MartinsPRB99} Perhaps 
this explains why RVB theories, which assume spin charge separation 
throughout the entire zone, incorrectly predict a cusp at 
$(\pi/2,\pi/2)$ in the spinon dispersion while still correctly
describing the overall dispersion on a qualitative 
level.\cite{LaughlinPRL97,WengPRB01}
	
The mean field treatment in a simple SDW picture applied to the Hubbard
model with up to third nearest neighbor hopping fails to describe the
data.\cite{TohyamaJPSJ00} The dispersion along $(\pi,0)$ to $(0,\pi)$ is
much too great for realistic parameters.  However, the dispersion does
contain a rounded nodal region.  If one can argue that one should
renormalize these parameters then it is possible to obtain a very good fit
to the experimental data as done by Nu\~{n}ez-Regueiro\cite{Regueiro}
where they also show that the magnitude of the overall dispersion
decreases with doping as has been seen in the high energy pseudogap.  
Alternatively, a diagrammatic expansion of the Hubbard model containing
$t^{\prime}$ and $t^{\prime\prime}$ as well as the QED$_{3}$ theory also
reproduce the observed dispersion.\cite{BruneEuroB00,Herbut02} Finally,
the SO(5) theory predicts a $|\cos k_{x}a-\cos k_{y}a|$ dispersion for the
insulator when the superconductor has simple $d$-wave
pairing.\cite{RabelloPRL98} However, in the projected SO(5), the observed
flattening of the node in the insulator is consistent with a flattened
dispersion observed near the node of the superconducting
gap.\cite{ZacherPRL00}
	
\subsection{E$_k$ isotropy of $(\pi/2,\pi/2)$}

\begin{figure}[tb]
        \centering
        \includegraphics[width=3in]{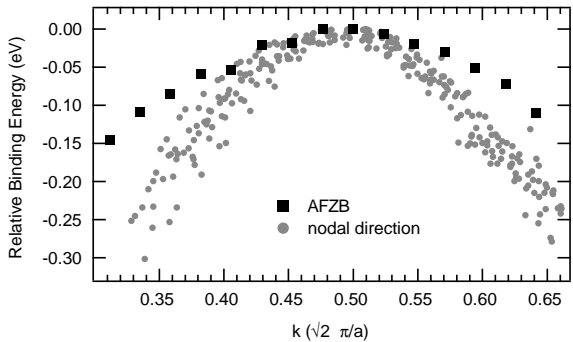}
        \vspace{0.1in}
\caption[Parameterizing the isotropy of $(\pi/2,\pi/2)$]
        {Anisotropy of the dispersion about $(\pi/2,\pi/2)$. The nodal 
direction data is reproduced from figure \ref{EkEphotsum}, including all 
13 photon energies at once, while the data along the 
antiferromagnetic zone boundary (AFZB) is reproduced from figure 
\ref{roundedBfits}. The dispersion in each case is found by taking the 
minimum of the second derivative of the respective EDCs.}
        \label{EkAnisotropy}
\end{figure}

      By combining our investigation of the nodal direction with that of
the rounding along the antiferromagnetic zone boundary we can get a
measure of the isotropy of the dispersion at $(\pi/2,\pi/2)$ shown in
figure \ref{EkAnisotropy}. The data demonstrates a clear anisotropy in the
two directions. However, it should be noted that in tracking the
dispersion from the EDC's that the total dispersion in both directions
does not exceed 350meV. It is only due to the improved momentum resolution
that we get a clear indication of the existence of some anisotropy in the
dispersion about $(\pi/2,\pi/2)$. It should be noted that the full
dispersion in the nodal direction is difficult to extract due to a loss
of spectral weight at (0,0) which has historically been attributed to
matrix elements. A similar lack of spectral weight near $(\pi,0)$ exists
in SCOC, but for CCOC the total dispersion along the antiferromagnetic
zone boundary is easily extracted (see figure \ref{SCOCvsCCOC}). The
resulting 350meV dispersion agrees with the estimates determined from
SCOC. We note that most models can easily account for a small degree of
anisotropy by introducing additional terms such as next nearest neighbor
hopping.

\section{Conclusions}

To summarize, we have studied the Mott-insulating cuprates:  
Sr$_2$CuO$_2$Cl$_2$, Ca$_2$CuO$_2$Cl$_2$, Ca$_2$CuO$_2$Br$_2$,
Sr$_2$Cu$_3$O$_4$Cl$_2$, Bi$_2$Sr$_2$ErCu$_2$O$_{8}$, and
Bi$_2$Sr$_2$DyCu$_2$O$_{8}$.  The lowest energy excitations measured by
ARPES are the same in these compounds as for the prototypical insulators:
Sr$_2$CuO$_2$Cl$_2$ and Ca$_2$CuO$_2$Cl$_2$.  Other studies on
La$_{2-x}$Sr$_{x}$CuO$_{4}$ and Nd$_{2}$CuO$_{4}$ have similar
results.\cite{InoPRB00,ArmitageNCO,Yoshida01} This shows that the low
energy excitations are indeed independent of the apical atom, and that the
results from Sr$_2$CuO$_2$Cl$_2$ and Ca$_2$CuO$_2$Cl$_2$ are truly
representative of a half-filled CuO$_{2}$ plane.  This justifies the
connection between the high energy pseudogap observed in underdoped Bi2212
and the $d$-wave-like modulation seen in
ACOC.\cite{RonningScience98,LaughlinPRL97} Our photon energy dependent
study enables us to extract the intrinsic spectral function which shows
that the dispersion is indeed independent of photon energy, and that
despite strong variations due to the matrix elements an asymmetry exists
in the n(${\bf k}$) about the line from $(\pi,0)$ to $(0,\pi)$.  We find
that the dispersion along this contour does not fit a simple $d$-wave
dispersion, but is flattened near $(\pi/2,\pi/2)$ consistent with a
similar rounding observed in underdoped Bi2212
samples.\cite{LoeserScience96,HarrisPRB96,MesotPRL99} Numerical
calculations on the $t$-$t^{\prime}$-$t^{\prime\prime}$-$J$ model which
match all aspects of the insulator dispersion including the rounded nodal
region, show that the antiferromagnetic correlations are destroyed near
$(\pi,0)$, indicating that a spin liquid picture may be more appropriate
for interpreting the ARPES data.\cite{TohyamaJPSJ00,MartinsPRB99} However,
generalized $t$-$J$ models fail to reproduce the observed asymmetric
intensity profile which is reproduced in the Hubbard
model.\cite{DagottoRMP94,EskesPRB96} Possibly the Hubbard model including
next nearest neighbor hopping terms\cite{BruneEuroB00} is necessary to
capture all the physics of a single hole in the half-filled CuO$_{2}$
plane.

\section{Acknowledgments}

We thank C. D\"urr, M.S. Golden, J. Fink, and T. Yoshida for very open 
and stimulating discussions.  This work was performed at Stanford 
Synchrotron Radiation Laboratory, a national user facility operated by 
Stanford University on behalf of the US Department of Energy, Office 
of Basic Energy Science.  The division of Chemical Sciences and 
Material Sciences also supported this work as did the ONR grant 
N00014-98-1-0195 and NSF grant DMR-0071897.


\begin{thebibliography}{99}
 
\bibitem[\ast]{fr}  {\it Present Address}: Dept. of Physics,
University of
Toronto, 60 St. George St, Toronto, ON, M5S 1A7, Canada
 
\bibitem[\dag]{cyk}  {\it Present Address}: Dept. of Physics, Yonsei
University, Seoul, Korea
 
\bibitem[\S]{npa}  {\it Present Address}: Dept. of Physics and Astronomy,
UCLA, Los Angeles, CA
\bibitem[\ddag]{dlf}  {\it Present Address}: Dept. of Physics and
Astronomy,
University of British Columbia, Vancouver, Canada
 
\medskip
%%%%%%%%%%%%%%%% Refs For Intro %%%%%%%%%%%%%%%%%%
\bibitem{Mott49}  N.F. Mott, Proc. Phys. Soc. A {\bf 62} 416 (1949)
\bibitem{InoPRB00} A. Ino {\it et al.}, Phys. Rev. B {\bf 62} 4137
(2000) %creating in gap states?
\bibitem{ArmitageNCO} N.P. Armitage {\it et al.}, Phys. Rev. Lett. 
{\bf 88}, 257001 (2002) %NCO data and doping dependence
\bibitem{HarimaPRB01} N. Harima {\it et al.}, Phys. Rev. B {\bf 64} 220507
(2001) %core level shifts in NCCO
\bibitem{SchabelPRB98} M.C. Schabel {\it et al.}, Phys. Rev. B {\bf 57}
6090 (1998)
\bibitem{Grande75} B. Grande and H. M\"uller-Buschbaum, Z. Anorg. Allg.
Chem. {\bf 417} 68 (1975)
\bibitem{LynchOlsonBook} D.W. Lynch and C.G. Olson, {\it Photoemission
Studies of High-Temperature Superconductors} (Cambridge University
Press, Cambridge, 1999)
\bibitem{ZhangRice88} F.C. Zhang and T.M. Rice, Phys. Rev. B {\bf 37},
3759 (1988)
\bibitem{HaffnerPRB00} S. Haffner {\it et al.}, Phys.  Rev.  B {\bf
61}, 14378 (2000)
\bibitem{DurrPRB00} C. D\"urr {\it et al.}, Phys. Rev. B {\bf 63},
014505 (2000)
\bibitem{HaffnerPRB01} S. Haffner {\it et al.},  Phys. Rev. B {\bf 63},
212501 (2001) %''No well defined Remnant Fermi Surface''
\bibitem{RonningScience98} F. Ronning {\it et al.}, Science {\bf 282}, 
2067 (1998).
\bibitem{LaughlinPRL97} R.B. Laughlin, Phys. Rev. Lett. {\bf 79}, 1726
(1997) %explains High E pseudogap
\bibitem{WhitePRB96} P.J. White {\it et al.}, Phys. Rev. B {\bf 54},
R15669 (1996)
\bibitem{WengPRB01} Z.Y. Weng {\it et al.}, Phys. Rev. B {\bf 63},
075102 (2001) %phase string of one hole in Mott AF.
\bibitem{RabelloPRL98} S. Rabello {\it et al.}, Phys. Rev. Lett. {\bf
80}, 3586 (1998) %SO(5)
\bibitem{LancePRB90} L. L. Miller {\it et al.}, Phys. Rev. B {\bf 41},
1921 (1990) %synthesis and structure of Sr2CuO2Cl2
\bibitem{KitajimaJPCM99} T. Kitajima {\it et al.}, J. Phys. Condens.
Matter {\bf 11}, 3169 (1999)
\bibitem{WellsPRL95} B.O. Wells {\it et al.}, Phys.  Rev.  Lett.
{\bf 74}, 964, (1995)
\bibitem{KimPRL98} C. Kim {\it et al.}, Phys. Rev. Lett. {\bf 80}, 4245,
(1998)
\bibitem{MattheissPRB90} L.F. Mattheiss, Phy. Rev. B {\bf 42}, 354
(1990) %band structure of CCOC and CCOB
\bibitem{MarshallPRL96} D.S. Marshall {\it et al.}, Phys. Rev. Lett. {\bf
76} 4841 (1996)
\bibitem{ShenPhysRep95} Z.-X. Shen and D. S. Dessau, Phys. Rep.
{\bf 253}, 1 (1995)
\bibitem{GoldenPRL97} M.S. Golden {\it et al.}, Phys.  Rev. Lett.
{\bf 78}, 4107 (1997) % Ba2Cu3O4Cl2 ``A tale of 2 singlets''
\bibitem{ShmelzPRB98} H.C. Schmelz {\it et al.}, Phys. Rev. B {\bf 57},
10936 (1998) % Ba2Cu3O4Cl2 second ARPES paper from Fink group
\bibitem{PothuizenPRL97} J.J.M. Pothuizen {\it et al.}, Phys. Rev.
Lett. {\bf 78}, 717, (1997)
\bibitem{LaRosaPRB97} S. LaRosa {\it et al.}, Phys.  Rev.  B
{\bf 56}, R525, (1997)
\bibitem{DagottoRMP94} E. Dagotto, Rev. Mod. Phys. {\bf 66}, 763, (1994)
\bibitem{StringResonance} E. Dagotto {\it et al.}, Phys. Rev. B
{\bf 41}, 9049 (1990); Z. Liu and E. Manousakis, Phys. Rev. B {\bf
45}, 2425 (1992) %string resonances
\bibitem{EskesPRB96} H. Eskes and R. Eder, Phys. Rev. B {\bf 54},
14226 (1996) %''pseudo'' Fermi surface
\bibitem{RanderiaPRL95} M. Randeria {\it et al.}, Phys. Rev. Lett. {\bf 
74}, 4951 (1995)
\bibitem{pseudogaps} For a review on ARPES of cuprates including a 
discussion of the pseudogaps see: A. Damascelli,  Z.-X. Shen, and Z. 
Hussain, To appear in Rev. Mod. Phys.  
\bibitem{LoeserScience96} A.G. Loeser {\it et 
al.}, Science {\bf 273}, 325 (1996)
\bibitem{HarrisPRB96} J.M. Harris {\it et al.}, Phys. Rev. B {\bf
54}, R15665 (1996)
\bibitem{MesotPRL99} J. Mesot {\it et al.}, Phys. Rev. Lett. {\bf 83},
840 (1999)
\bibitem{KohsakaThesis} Y. Kohsaka, Masters Thesis, University of
Tokyo, 2001
\bibitem{KaminskiPRL00} A. Kaminski {\it et al.}, Phys. Rev. Lett.
{\bf 84}, 1788 (2000) %''quasiparticles in SC state of Bi2212''
\bibitem{KimPRB02} C. Kim {\it et al.}, Phys. Rev. B {\bf 65}, 
174516 (2002) %anomalous T dep. of SCOC.
\bibitem{Yoshida01} T. Yoshida {\it et al.}, 
http://xxx.lanl.gov/cond-mat/0206469
%new LCO data in second zone including x dep.
\bibitem{KastnerRMP98} M.A. Kastner, {\it et al.}, Rev. Mod. Phys. {\bf 
70}, 897 (1998) %Mag, transport, and optical props. of monolayer cuprates
\bibitem{HiroiNature94} Z. Hiroi, N. Kobayashi, and M. Takano, Nature, 
{\bf 371}, 139 (1994); Z. Hiroi, N. Kobayashi, and M. Takano, Physica C, 
{\bf 266}, 191(1996) %NaCCOC crystal growth
\bibitem{AlMamouriNature94} M. Al-Mamouri {\it et al.}, Nature {\bf
369}, 382 (1994) %discovery of first superconducting oxyhalide: SCOF
\bibitem{NovikovPRB95} D.L. Novikov, A.J. Freeman, and J.D. 
Jorgensen, Phys. Rev. B {\bf 51}, 6675 (1995).
\bibitem{Bulut} N. Bulut, D.J. 
Scalapino, and S.R. White, Phys. 
Rev. Lett. {\bf 73} 748 (1994).
\bibitem{LemaPRB97} F. Lema and A.A. Aligia, Phys. Rev. B {\bf 55},
14092 (1997)
\bibitem{TohyamaReview} T. Tohyama and S. Maekawa, Supercond. Sci.
Tech. {\bf 13} R17 (2000) and references therein
%Theoretical review of High Tc ARPES
\bibitem{TohyamaJPSJ00} T. Tohyama {\it et al.}, J. Phys. Soc.
Japan {\bf 69}, 9 (2000)
% ``spin liquid around a doped hole in insulating cuprates''
\bibitem{MartinsPRB99} G.B. Martins, R. Eder, and E. Dagotto, Phys. Rev.
B {\bf 60}, R3716 (1999) %Similar to TohyamaJPSJ00
\bibitem{Regueiro} M.D Nu\~{n}ez-Regueiro, Euro. Phys. J. B {\bf
10} 197 (1999)
\bibitem{BruneEuroB00} P. Brune and A.P. Kampf, Eur. Phys. J. B {\bf
18}, 241 (2000) %diagramtic expansion of Hubbard-t'-t'' model
\bibitem{Herbut02} I.F. Herbut, http://xxx.lanl.gov/cond-mat/0202491
\bibitem{ZacherPRL00} M.G. Zacher {\it et al.}, Phys. Rev. Lett. {\bf
85}, 824 (2000) %SO(5): round Bi2212 gives rounded insulator

\end{thebibliography}
\end{document}